# How learners produce data from text in classifying clickbait

Nicholas J. Horton (nhorton@amherst.edu), Jie Chao, Phebe Palmer, and William Finzer

## Abstract

Text provides a compelling example of unstructured data that can be used to motivate and explore classification problems. Challenges arise regarding the representation of features of text and student linkage between text representations as character strings and identification of features that embed connections with underlying phenomena. In order to observe how students reason with text data in scenarios designed to elicit certain aspects of the domain, we employed a task-based interview method using a structured protocol with six pairs of undergraduate students. Our goal was to shed light on students' understanding of text as data using a motivating task to classify headlines as "clickbait" or "news". Three types of features (function, content, and form) surfaced, the majority from the first scenario. Our analysis of the interviews indicates that this sequence of activities engaged the participants in thinking at both the human-perception level and the computer-extraction level and conceptualizing connections between them.

Running head: Producing data from text

Keywords: Teaching Statistics, Classification, Feature Engineering, Text Analytics, Data Science

# How learners produce data from text in classifying clickbait

Abstract

Text provides a compelling example of unstructured data that can be used to motivate and explore classification problems. Challenges arise regarding the representation of features of text and student linkage between text representations as character strings and identification of features that embed connections with underlying phenomena. In order to observe how students reason with text data in scenarios designed to elicit certain aspects of the domain, we employed a task-based interview method using a structured protocol with six pairs of undergraduate students. Our goal was to shed light on students' understanding of text as data using a motivating task to classify headlines as "clickbait" or "news". Three types of features (function, content, and form) surfaced, the majority from the first scenario. Our analysis of the interviews indicates that this sequence of activities engaged the participants in thinking at both the human-perception level and the computer-extraction level and conceptualizing connections between them.

## Introduction

The National Academies of Science, Engineering, and Medicine consensus report on "Data Science for Undergraduates: Opportunities and Options" noted that:

> All data scientists need to learn how to tackle questions with real data. It is insufficient for them to be handed a "canned" data set and be told to analyze it using the methods that they are studying. Such an approach will not necessarily prepare them to solve more realistic and complex problems taken out of context, especially those involving large, unstructured data. Instead, they need repeated practice with the entire cycle beginning with ill-posed questions and "messy-data" (NASEM, 2018, page 2-6).

Text provides an important example of unstructured data that has traditionally been nearly absent from the statistics curriculum in secondary schools and early post-secondary education), despite the importance of the written word in all aspects of education and society.

Gentzkow, Kelly, and Taddy (2019) note that text differs from other typical types of data given its inherent high dimensionality and need for different statistical methods for analysis. How can such "messy-data" be represented? Konold, Finzer, and Kreetong (2017), who defined a "case" as "the physical record of one repetition of a repeatable observational process", noted that data are typically recorded in table format. This immediately raises important questions regarding text as data, since text is obviously stored as data on a computer, but its underlying phenomena (meaning) is of interest, rather than the specific encodings.

Inferring underlying meanings is the province of natural language processing, an area that has historically developed within the artificial intelligence community. However, classification and predictive analytics (Breiman, 2001) within textual analysis provide a natural bridge to statistics and data science. For example, an analyst might be interested in a model for classifying a given social media post as either misinformation or a legitimate news item. Such models are increasingly important in a world full of contradictory information sources. Classification via 2x2 tables, logistic regression, and decision trees have become more common components of introductory statistics and data science education (Zieffler et al, 2021; Baumer et al, 2021; Hazzan and Mike, 2022) and thus provide an earlier entry point for student analysis of simple text-based features.

Limited prior pedagogical research has been undertaken in this relatively new area of text analysis in statistics and data science education. Many efforts have been made to use text analytics to automate assessments and identify new insights, but few on student comprehension of the process behind producing data from text (Gentzkow, Kelly, and Taddy, 2019; Jiang et al., 2022).

**Features and feature engineering**

A key aspect of text analysis relates to how features or attributes are extracted from an underlying text string. Feature engineering is a broad process, but it includes the conversion of raw text into feature values that encode information about the presence of specific attributes (the phenomena of interest) in the text. It includes a range of activities including feature extraction, feature selection, and dimensionality reduction. Some commonly used features include unigrams (individual words), n-grams (combinations of words), punctuation, length, and the presence of certain keywords (see Silge and Robinson, 2017).

Prior work has explored some aspects of feature creation. Fiebrink (2020) discusses aspects of pedagogy for feature extraction in machine learning. There have been a few reports on students' inability to discover new insights in the context of social media analytics (e.g., Goh & Sun, 2015); however, minimal research has focused on how to help students develop such skills. This is in contrast to topics such as decision trees for classification (e.g., CATALST Group, 2009; Engel, Erickson, and Martignon, 2019) that have received more prior attention.

Hardy, Dixon, and Hsi (2020, Table 1) provide a framework for data production, including how data are produced by "instruments" (e.g., raw sensors) that, like text, often feature high dimensionality, through disciplinary practices that embody disciplinary knowledge, and for human purposes. We see the creation of features from text strings to be an example of such data production.

Bishop (2006) notes that a feature includes useful discriminatory information for a classification problem. It represents a measurable property of a phenomenon. In statistics we might describe it as a "predictor" or "covariate" within a modeling framework. Features can be human-perceivable, able to be extracted by a computer, or increasingly both.

## Design

### Learning Objectives

Our research explores questions regarding beginning and intermediate student understandings of text as data. Our ultimate goal was to support students in developing an understanding of features in text classification models. Students have much prior knowledge that can be drawn upon. For example, they can read text and comprehend the meanings. They also know that computers store sequences of characters that make up words, phrases, sentences, and paragraphs. But how can human understanding be represented as low-level features using computers? That is the key idea that we want to help students understand.

Our design was guided by the Cultural-Historical Activity Theory (CHAT, Engeström, Miettinen, & Punamäki, 1999), which provided a powerful lens to understand how learning occurs. CHAT views human activities as goal-oriented systems that comprise dynamic interactions among subject, object, mediating artifacts, community, rules, and division of labor. As a subject attempts to achieve an object (or goal), his or her actions are mediated by artifacts, supported by the community, and constrained by the rules and division of labor in the community. From the CHAT perspective, learning is a by-product of the activity system. In trying to achieve the object, the subject must acquire, use, and adapt mediating artifacts, which include both tools and mental instruments such as concepts, schemas, and procedures. The development of mental instruments is equivalent to the conventional notion of learning. The characteristics of other system components (i.e., rules, community, and division of labor) shape the development process and outcomes. Thus, for the same object, the subject may develop and use different mental instruments because certain activity system components are different. For instance, people adjust their communication styles based on the audiences (e.g., explaining statistics to a college student versus an elementary student). The different characteristics of the audiences elicit and prioritize different sets of knowledge elements, which can be constructed into related but distinct understandings.

**Design Conjecture**

Our overall conjecture was that different types of information-processing agents (oneself, others, and computers) present different but overlapping sets of affordances and constraints, which can draw students' attention to different types of features and help them make connections between them.

This overarching design is embodied as follows. Students were first presented with a professional context and the need to identify "clickbait" articles. We define "clickbait" as headlines that appear in emails or web pages that are designed primarily to get you to click so that you will view advertising on a site (EAVI, 2022). Clickbait, described by EAVI as a "low impact" form of misinformation, provides a tangible way to approach broader issues regarding disinformation. It also has the advantage of being familiar to students and of interest as a motivating example[1].

Students were then asked to identify features useful for sorting headlines into clickbait and news in three different but related scenarios. Each scenario introduced a different information-processing agent: 1) students themselves, 2) other human analysts, and finally 3) computers. The characteristics of these information-processing agents can draw students' attention to different types of features: some are human-perceivable and others are computer-extractable. The relationships among these information-processing agents highlight the connections between the two types of features.

**Scenario 1. Two raters labeling text and seeking agreement**

The first scenario required students to collaboratively label a few headlines as clickbait or news and then reflect on how they made their decisions.

Sample clickbait headlines (taken from sites known to host clickbait) include:

1. This Freestyle Crossword Is For Everyone Who Loves Pop Culture
2. How Over Winter Are You
3. 32 Products Every Elephant Lover Needs In Their Home

Non-clickbait headlines (headlines from traditional news sources):
1. WHO: Polio reemerging in Africa
2. BP: New cap on Gulf of Mexico oil well in place

---

[1] Others have explored ways to identify clickbait. Chakraborty et al (2016) developed a classification model for clickbait; similar methods have been improved since that time (Mari, 2022). Scott (2021) undertook a corpus analysis that explores aspects of clickbait using an information gap approach. Horton et al (2022) describe ways to bring text analysis into the classroom using a related task (classifying emails as spam or legitimate messages).

    3. US presidential candidate Obama speaks in Berlin, Germany

Students were asked to discuss their ideas and reach a consensus among themselves. This requirement created the opportunity for students to explain their initial ideas by citing certain features, to discuss why these features indicate the labels, and to make decisions based on the relationships among these features. Because the first scenario was focused on labeling text by hand and human decision-making, students are likely to identify human-perceivable features. Their informal or formal knowledge of the domain may influence their ideas. For example, students may explain that clickbait is designed to make people curious and click on the link and identify "making people curious" as a key feature, which can only be perceived by humans.

**Scenario 2. Identifying features to train others**

The second scenario introduced the need to train new recruits to a hypothetical company to detect clickbait. In preparation for training others, students were given a larger set of clickbait and non-clickbait headlines[2]. The students were asked to identify features that can distinguish between clickbait and news headlines (note that the term "feature" was used in the task description and prompts in a common way without being strictly defined as we have done above). The information-processing agent in this scenario is other people who have the same perceptual and cognitive abilities as students themselves do. However, they may vary in how they interpret features, especially those subjective features such as "making people curious".

To improve the reliability of other people's work, we suspect that students may draw on their knowledge of the English language to formulate rules or give examples. For instance, curiosity-eliciting headlines are typically phrased as questions or extreme situations. These rules are more objective and get closer to the kinds of features extractable by computers. Students might also notice some objective features that are good indicators even though they cannot explain how those features work. For example, clickbait headlines tend to start with numbers, a feature that has a deeper psychological explanation. Regardless of the underlying reasons, these objective features are useful for training others to reliably sort the headlines.

---

[2] Instructors interested in utilizing this example in their courses can find a CODAP document at https://codap.concord.org/releases/latest/static/dg/en/cert/index.html#shared=158719. Other approaches to teaching classification of text using a variety of computational tools can be found in Horton et al (2022).

**Scenario 3. Automating text classification**

The third scenario introduced the need to automate the clickbait detection work to classify a large number of headlines. Three prompts were designed to guide students to make connections between human-perceivable and computer-extractable features. The first prompt asked students to review the features they previously identified and pick out the ones that can be detected by computers. Students are typically familiar with common text processing tools such as word search and character count. We anticipated that they should be able to identify features like "starting with a number" or "including a certain topic word". They may also start to operationalize the rules that they formulated in the previous scenario to features detectable by these tools. For example, questions, which are often used in clickbait, can be further operationalized as the presence of question words and/or question marks.

The next prompt asked students to analyze the human-perceivable features (e.g., felt gimmicky, false advertisement, etc.) and identify indicative features detectable by computers. Students may draw on their knowledge of the English language and related domains (e.g., popular culture) to do the analysis. For example, clickbait often claims extremity, which can be indicated by the use of the superlative form of adjectives. Clickbait also frequently references popular cultures and celebrities, which can be enumerated in a list that a computer can use to look up.

The final prompt asked students to bring all computer-extractable features into one single list in preparation for the next part of the activity (only the feature creation component of the interview was analyzed in this paper).

**Research Questions**

The goal of this study was to evaluate the design and validate the underlying theory described above. To achieve this goal, we conducted an observational study to answer the questions:

1. What types of features do participants identify when completing a sequence of tasks designed to support them in recognizing both human-perceivable and computer-extractable features and the connections between the two?
2. What characteristics of the tasks appear to facilitate or impede participants' thinking?

## Methods

In order to observe how students reason with text data in scenarios designed to elicit certain aspects of the domain, we employed a task-based interview method with a structured protocol (Maher & Sigley, 2020). The project was approved by the Amherst College Institutional Review Board.

We piloted a performance task involving identifying the characteristics of "clickbait" headlines with six pairs of undergraduate students. Our methodology involved a task-based interview of approximately 60-90 minutes in duration. Each of the six pairs participated in a recorded Zoom meeting with screen sharing. Four of the pairs consisted of students who had taken an introductory statistics and an introductory data science course; the remaining two pairs had taken only an introductory statistics course. Notes and recordings were coded and analyzed to address key research questions.

The second author analyzed the participants' notes using the open coding method (Strauss, 1987). Codes were applied for unique features or rules identified by the participants. The codes were then categorized by their targets (clickbait or news), dimensions (i.e., impacts on readers, topics, styles), and feature types (i.e., words, linguistic structures, etc.). These codes and categories were further mapped onto the five prompts in the first part of the interview to show how different ideas emerged as responses to the prompts designed to elicit certain ways of thinking.

To facilitate interpretation of the task and the problem, we have included excerpts from the two interviews of science majors who had only taken an introductory statistics course. This included a pair of sophomore students (we refer to them as "Leela" and "Kevin") and a pair of junior students (we refer to them as "Emily" and "Jenna").

## Results

**Responses to Scenario 1.** The interview started by inviting the participants to classify five headlines, where three of them were clickbait and the other two were news. As an example, one pair had the following interaction:

> (Participants looking at the headline "That $1 Trick Will Help You To Get Amazing Even Eyebrows")
> **Leela**: Oh, clickbait. I've seen this exact one.
> **Kevin**: Oh, really?
> **Leela**: I'm pretty sure, something like…
> **Kevin**: Gotcha. Did you click on it?

> **Leela**: No, because then that gives them money and I'm against that.

After they completed classifying the first five headlines, they were presented with the five headlines and true label (clickbait vs. news) altogether and prompted to explain how they made decisions about them. Here are examples from two of the pairs:

> **Emily**: We assessed the plausibility of the headline using our knowledge of science and current events. We gauged whether or not the headline seemed like it was trying to draw the reader into the article.
> **Jenna**: I feel like the ones that were clickbait used certain words, I don't know how to describe them, that were "dramatic" or "gimmicky" to indicate that the article was clickbait.
> **Leela**: So I guess like, one of the reasons was if the title seemed like exciting or boring for me at least.
> **Kevin**: I think if it's a more informational or you know like informative title, it tends to be a news headline. Like "scientists in Germany draft the Neanderthal" or "no union for Fedex home drivers court rules" that's kind of I guess informative right off the bat and everything else seems to be more like just trying to pique your curiosity like "what's your favorite food" or "17 of you know photos that offer that rare look" and then "this one dollar trick will help you get amazing even eyebrows". It seems to be more like trying to pique your curiosity, not so much like an informative kind of thing like the first one, the other two that I mentioned.
> **Kevin**: What do you think, Leela?
> **Leela**: Um yeah, I totally agree.
> **Kevin**: Gotcha.

**Results from all six pairs.** Analysis of all six pairs' responses showed that the participants identified a wide range of features along three dimensions of clickbait: *function*, *content*, and *form*.

Along the *function* dimension, some participants described clickbait as "gives them money," "advertisement," "draw the reader into the article," and "encourage readers to click on it," which are the end goal of clickbait. Many participants further identified the feelings or impressions elicited by clickbait, intentionally or unintentionally: "intended to elicit certain feelings", "pique curiosity", "interesting", "exciting", "boring," "dramatic", "vague", and "gimmicky."

In terms of *content*, some participants pointed out that the content of clickbait are typically "everyday problems" and "inconsequential".

With regard to *form*, a few participants also paid attention to the form of clickbait and discerned "certain words" and "individualized wording and language" in clickbait. The news headlines, by contrast, were discussed much less explicitly by the participants. Only two of the six pairs pointed out that news headlines typically presented new, novel, and informational/informative content. "knowledge of science and current events"

While all three types of features (*function*, *content*, and *form*) surfaced at this stage, the majority of the features were about the *function* and *content* of clickbait, which require human interpretation and perception. For instance, only humans can perceive a headline as interesting, notice dramatic words, and feel the urge to click to see more.

Analysis of selected pairs' conversations showed that the participants exchanged most ideas around the *function* and *content* of the headlines. Jenna, for example, noticed that "clickbait used certain words," but instead of listing those words, she shifted to describe the impressions produced by those words to Emily. Similarly, Kevin pointed out news headlines were informational or informative and read through a few headlines as examples. In both cases, the participants communicated with the implicit assumption that their partners were capable of perceiving and understanding the features they identified. Their partners' immediate feedback also reinforced their implicit assumption.

**Responses to Scenario 2.** Immediately following the first prompt, the interviewer presented a slightly different requirement for the same classification task. Instead of performing the classification task themselves, the participants were asked to come up with a set of instructions for training other employees to perform the task. A new set of 20 headlines (10 clickbait and 10 news) were presented on the same page. Specifically, the participants were asked to first observe and "take note of any features of the headlines" that distinguish clickbait and news and then get together to discuss "the features". Note that the word "feature" was used here informally without being defined as extractable features in text mining.

Analysis of all pairs' notes showed that, in response to the second scenario, the participants identified more features for both clickbait and news headlines.

They discovered that the content of clickbait was characterized by attempts to generalize and direction toward the readers. They also identified many characteristics of clickbait's *form*: using "fluffy" adjectives, using second-person point of view with words like "you" and "your", referring to things directly, asking questions, suggesting a number of items for a phenomenon or problem, and containing or starting with numbers, and containing words like "everyone," "this," "every," and "reasons".

The participants identified many more features of news headlines than they did in the first scenario. They described the function of news headlines as "to inform" or "informational". With regard to *content*, they noted that news headlines presented consequential and factual content (e.g., death, abortion, natural disaster) related to world events, governments, and specific locations (e.g., words like "China" and Germany")". In terms of *form*, the participants found that news headlines were typically descriptive, summarized the content well, used a third-person point of view, and followed a formulaic sentence structure.

***Explaining and giving examples.*** Many computer-detectable features emerged in this stage as the participants explained their ideas and gave examples. For instance, two pairs mentioned formulaic sentence structure and gave specific examples like "who, what, when, where, and maybe why," and "subject + verb". One pair noted the common characteristic of localization was directly associated with words like "China" and "Germany".

They would identify a *form* feature and point out its textual evidence. Often, the participants spontaneously made connections between the *form* and the *function* of clickbait. For instance, one pair wrote, "if the headline specifically targets a certain group of people, using words like 'you' and 'your' then it's most likely clickbait". Analysis of selected participants' interview data showed that the participants reasoned about how the *form* and *content* of clickbait or news headlines generate the intended effects, or their *functions*. This reasoning gave rise to more identified features and the connections among them.

For instance, continuing with Kevin's and Leela's conversation, Kevin noticed the *form* being "second-person view" and "personalized", and then linked it to the *function* "pique the reader's interest".

> **Kevin**: I noticed for the clickbait websites, they tend to be kind of referring to like second-person view. They're kind of personalized and like I said it kind of seems to be trying to pique the reader's interest…

He further explained the link between the *form* and the *function*: personalization can stimulate interest because people tend to be self-interested.

> **Kevin**: … so obviously like personalizing it or saying "you" or involving you maybe it's kind of like interesting since people can be self-interested at times. So they may be curious about, you know, what their favorite food is based on the zodiac signs …

In addition, he contrasted clickbait and news headlines and pointed out the "impersonal" but "informative" nature of news headlines:

> **Kevin**:… and then for the traditional news sites seems to be kind of the opposite they're not, they're impersonal. They're about, you know, things that happen in the world, kind of informative, anything but you unless you, you know you're actually involved in those news headlines.

Sometimes, the participants noticed a *form* feature but were not able to explain how it worked to produce the *function*. For example, one pair wrote, "The clickbait articles tend to use titles that include words that refer to people or things directly (such as 'this', 'every', 'you', etc.)"

Some links they noted were implicit. For example, one pair pointed out, "If the headline suggests a certain number of items to represent some phenomenon or solve a daily problem, then it's most likely clickbait." In this statement, the participants' focus was on the human-perceivable features "a certain number of items to represent...", yet the containing-numbers feature was clearly visible.

**Responses to Scenario 3**. The participants were further prompted to determine which previously identified features could be detected by a computer, to focus on those human-perceivable features, identify indicative structural elements of the text, design rules for a computer to detect them, and to create a list of as many attributes of the headlines for identifying clickbait using a computer.

Jenna and Emily reasoned about the structures of clickbait and news headlines and identified the key words that make the clickbait either personal or exaggerating:

> **Jenna**: I feel like the clickbait ones: the way they structure them, the headline is talking to you rather than just stating a fact. Whereas the news ones are less personal, they are like stating "someone did this or something did this", a subject and an action, while the clickbait ones used "you" and "we" a lot. They include numbers more often.
> **Emily**: Yeah. And there are questions, like "are you more", exaggerating words, fluffy adjectives "pretty pretty cute".

Leela and Kevin were more focused on the features themselves and specific computational methods to extract them:

> **Leela**: Well, for the location thing there could be like a whole list of preset locations and then the computer could like match the match those locations with the words in the title and then if none of them come up then maybe put it into like a potential clickbait pile.
> **Interviewer**: Yeah, that's a good one to include.
> **Kevin**: Well what about like, if an article title starts off with a number like instead of the, you know, the spelled out number it gives you the the numerical number

> and it starts off like that. Maybe that could be a way to you know isolate the those clickbait articles.
> **Leela**: Uh-hum. Or if there's like the word "you" or "your" that's also clickbait.
> **Kevin**: Yeah, I think that's fair.

More generally, the participants' responses were mostly drawn or inferred from their previous responses: searching for numbers and trigger words for point of view (e.g., "you", "we"), acronyms, location (names of countries), government-related, questions, hyperboles, and mythical/fantastical elements. A few participants also noted styling such as using upper case.

Kevin first brought up the vocabulary level in clickbait headlines.

> **Kevin**: So what about like I guess like simpler words, words that kind of are really easy to understand by most people, so like, you know like "itty-bitty".
> **Leela**: Right, yeah.
> **Kevin**: Or like—

Building on Kevin's idea, Leela recognized casualness and familiarity as an important characteristic of clickbait. This idea seemed to have prompted the pair to quickly identify a slew of features including using first or second person pronouns, certain adjectives and adverbs, and certain nouns that refer to pop culture.

> **Leela**: I guess the words that invoke more of a casual sense?
> **Kevin**: How would you, how would you define that though?
> **Leela**: Well I guess there was that like the second person and the first person like if they use anything besides [muffled] third person then that could trigger it or there could be words like like "pretty" and "cute" and—
> **Kevin**: Oh yeah, yeah, right.
> **Leela**: Or also words like "really" or "very".
> **Kevin**: Right, yeah that's a good one.
> **Kevin**: Oh right and then I was thinking if it has nouns that are just references to like pop culture or like [muffled] but like fantastical nouns maybe like you know "elf" or—
> **Leela**: Hogwarts!
> **Kevin**: Yeah, that kind of, that can automatically maybe put on the clickbait list.
> **Leela**: Uh-huh. Yeah.
> **Leela**: And then also like what you said before "reasons" and then maybe also like "products".
> **Kevin**: Yeah, yeah.

One of the pairs made the explicit linkage between these rules and the concept of an algorithm (a specific procedure to solve a problem or perform a computation).

> **Interviewer:** Can you think of any ways that those rules could be broken down into similar sorts of computer-based rules?
> **Kevin**: So like computer-based rules you mean like some kind of algorithm that's like you could create to really easily sort the list. Is that what you're talking about?
> **Interviewer**: Yeah.

Most of the participants converged on detecting specific words for three characteristics: colloquial, indicated by words like "cute", "pretty", "really", "very", and "reasons"; fictitious, indicated by words like "Hogwarts", "elf", and "hobbit"; and location-related, indicated by city or country names. One pair identified a slightly complex structure named "adjectives/descriptors used sequentially" but were not able to explain which characteristic was related to it. Another pair devised an algorithm that first labels parts of speech, applies prebuilt wordlists, and computes the proportion of presence of news or clickbait related words.

Finally, the participants were asked to create a list of as many attributes of the headlines for identifying clickbait using a computer. The participants' responses were largely drawn from their previous notes and focused on the computer-extractable features that they would work with in the following part of the interview.

## Discussion

In this study, we observed what features college students identify when presented with three different information-processing agents (themselves, other humans, and computers). In terms of our first research question, the results showed that the participants identified a variety of features along three dimensions of media: *function*, *content*, and *form*. The majority of the human-perceivable features associated with function and content emerged during the first scenario, in which the participants collaboratively labeled headlines and reflected on their thought processes. When working on the second scenario, the participants not only identified additional human-perceivable features but they made connections between these three dimensions. The results from the third scenario were generally taken from their previous responses, with additional focus on strategies to identify features (e.g., by detecting words which match those on a specified list).

In terms of our second research question, the groups appeared to be fairly facile with their explorations and seemed to be able to transition from one scenario to the next with relative ease. We suspect that students with less experience with data science and statistics might find these transitions more challenging.

These results were fairly consistent with our expectations. The first scenario required the participants to collaboratively label the headlines. The information-processing agents were two human raters themselves. The capability of the information-processing agent oriented the participants' communication and focused their attention to human-perceivable features. The second scenario oriented the participants to identify features for training other employees. They were not always able to explain how a *form* feature functioned. The third scenario built on students' prior knowledge about text processing tools.

Overall, our analysis of the interviews indicates that this sequence of activities engaged the participants in thinking at both the human-perception level and the computer-extraction level as well as the connections between them. Students seemed to communicate that a feature was "an indicator of the target class" and, in the third scenario, see how it might be extractable by computing tools. We believe that the interviews help better understand ways that undergraduate students (some novice and others more advanced) think about "text as data".

Connecting these observations back to CHAT (the cultural-historical activity theory), the information-processing agents appeared to map onto both the mediating artifacts component, specifically tools, and the division of labor component. As our analysis showed, the participants' work was oriented by the capabilities of information-processing agents (tools) and the roles of themselves and the information-processing agents (the division of labor). As a guiding framework for this study, CHAT was proven to be useful. Future research can be further guided by CHAT to explore other design conjectures. For instance, manipulating certain properties of one type of information-processing agent may support students in differentiating related but distinct concepts (e.g., trainees with different backgrounds or computing tools with varied text processing capabilities).

**Implications**

Considerable effort has been made to "understand" human language using natural language processing and AI (artificial intelligence). These methods involve aspects of linguistics, machine learning, and sophisticated deep learning models to extract insights. In recent years, huge advances have been made in such tools (see for example the implications of ChatGPT, a conversational bot that can respond to users' questions and generate well-formed essays [Hirsh-Pasek and Blinkoff, 2023]). These developments have considerable implications for data science education.

Prior work at the school level has explored ways to prepare students for developments in this area. A number of relevant definitions and topics are included in the draft "Big Idea Progression" chart created by the Artificial Intelligence for K-12 project (AI4K12.org, 2022). We expand on these in detail because we believe that they are perhaps less familiar to many in the statistics and data science education community, are highly relevant, and provide a promising framework for broader consideration of how to prepare students to deal with more complex data sources and the role of AI in their lives.

Big Idea #1, entitled "Perception", describes a number of learning outcomes and enduring understandings relevant to how computers perceive the world, how perception is the extraction of meaning from this sensory information, and how the transformation from "signal to meaning takes place in stages, with increasingly abstract features and higher level knowledge applied at each stage." The concepts that comprise big idea #1 include:

1. *sensing* (by living things, computer sensors, and digital encoding),
2. *processing* (sensing vs. perception, feature extraction, abstraction) and
3. *domain knowledge*.

Example learning outcomes for "feature extraction" include:

1. "Give examples of features one would look for if one wanted to recognize a certain class of objects (e.g. cats) in an image (K-2),
2. "Illustrate how face detection works by extracting facial features" (3-5),
3. "Illustrate the concept of feature extraction from images by simulating an edge detector" (6-8),
4. "Explain how features are extracted from waveforms and images" (9-12).

It's worth noting that this progression chart focuses on examples using audio, image, and videos rather than text but we believe that the same concepts apply.

Big Idea #2, entitled "Representations and reasoning", introduces concepts about how representations support reasoning, that two major types of representations are symbolic and numerical, that "knowing" something requires representation and reasoning, and how "intelligent agents" employ a cycle of *sensing*, *deliberating*, and *acting*. For 3-5 graders, reasoning is explored through a learning outcome "categorize problems as either classification problems or search problems", with the former assigning observations to one of a predetermined set of classes. For 6-8 graders, representation is explored using feature vectors, with word embeddings noted as a key part of natural language processing using neural networks. In grades 6-8, more practice is given categorizing problems as classification, prediction, combinatorial search, or sequential decision problems as we have discussed in this paper. Other prior examples in the statistical education literature include classification of SPAM (CATALST group, 2009; Horton et al, 2022) and predicting tomorrow's temperature. The AI for K12 project suggests that students in later grades explore classification using richer and more complex data sources (e.g., images).

Big Idea #3, entitled "Learning", describes a number of learning outcomes and enduring understandings relating to how computers infer meaning from data. These include the *nature of learning* (humans vs. machines, finding patterns in data, training a model, constructing vs. using a reasoner, adjusting internal representations, and learning from experience), *neural networks* (structure of a neural network and weight adjustment), and *datasets* (feature sets, large datasets, and bias). Example learning outcomes for "feature sets" include:

1. "Create a labeled dataset with explicit features to illustrate how computers can learn to classify things like foods, movies, or toys" (K-2),
2. "Create a labeled dataset with explicit features of several types and use a machine learning tool to train a classifier on this data" (3-5),
3. "Create a dataset for training a decision tree classifier and explore the impact that different feature encodings have on the decision tree" (6-8), and
4. "Compare two real world datasets in terms of the features they comprise and how those features are encoded" (9-12).

The enduring understanding associated with the grade 9-12 LO is that "Humans decide which features to include in a dataset and how to encode them. This can have consequences for machine learning algorithms trained on these datasets."

We were struck with how the big ideas from the AI4K12 project, particularly "Representation" and "Perception", dovetail well with our research focus on extracting features from text. Our goal was to design a sequence of tasks that would lead to student mastery of this material. In this paper, we explored ways that instructors and curriculum developers can begin to expose data science students to text data. Such research is a necessary precursor to a research-based curriculum design. More specifically, we proposed to explore how students identify features in text and, in future research, use these features to classify text. We believe that this work is relevant for both school-based and university-based curriculum given the increasing growth of school-based data science (Biehler et al, 2022; NASEM, 2022).

**Closing thoughts**

Text analytics has the potential to help students address many ill-posed or ambiguous questions using data that they are familiar with through reading. As a relatively new field, text analytics has the potential to demonstrate how to extract meaning from new data sources. These types of data are attractive as they can be used to engage a number of motivating examples relevant to students and society. Teaching text analytics may also help to better integrate statistics and data science education with other subjects in primary and secondary education. Text provides a clear example of more complicated yet inherently interesting data sources that can be used to build a foundation in data acumen (NASEM, 2018).

Clickbait classification provides an accessible entry point to the development of students' data skills. In a world full of data it is critical that we prepare students to make sense of false information. As we consider ways to broaden statistics and data science education to incorporate more classification and predictive analytics, it will be important to identify approaches and techniques that effectively build student data skills.

## Acknowledgements

This material is based upon work supported by the National Science Foundation under Grant No. DRL-1949110. Support was also provided to Nicholas Horton through the Robert F. Tinker Fellowship at the Concord Consortium. Any opinions, findings, and conclusions or recommendations expressed in this material are those of the author(s) and do not necessarily reflect the views of the National Science Foundation or the Concord Consortium. We are appreciative of useful comments on an earlier draft from a number of SRTL 12 participants as well as the anonymous reviewers on an earlier draft. We have no conflicts of interest.